# Superconductivity and Rattling under High Pressure in the β-Pyrochlore Oxide RbOs$_2$O$_6$

Nao TAKESHITA*, Hiroki OGUSU[1], Jun-ichi YAMAURA[1], Yoshihiko OKAMOTO[1], and Zenji HIROI[1]

*Electronics and Photonics Research Institute, National Institute of Advanced Industrial Science and Technology, 1-1-1 Higashi Tsukuba, Ibaraki 305-8562, Japan*

[1]*Institute for Solid State Physics, University of Tokyo, 5-1-5 Kashiwanoha, Kashiwa, Chiba 277-8581, Japan*

*E-mail address: nao@takeshita.org

Rattling-induced superconductivity in the β-pyrochlore oxide RbOs$_2$O$_6$ is investigated under high pressures up to 6 GPa. Resistivity measurements in a high-quality single crystal show that the superconducting transition temperature $T_c$ increases gradually from 6.3 K at ambient pressure to 8.8 K at 3.5 GPa, surprisingly remains almost constant at 8.8 ± 0.1 K in a wide pressure range between 3.5 ($P_o$) and 4.8 GPa, and suddenly drops to 6.3 K at $P_s$ = 4.9 GPa, followed by a gradual decrease with further pressure increase. Two anomalies in the temperature dependence of the normal-state resistivity are observed at $P_o < P < P_s$ and $P > P_s$, revealing the presence of two high-pressure phases corresponding to the changes in $T_c$. The rattling of the Rb ion inside a cage made of Os and O atoms may be slightly and seriously modified in these high-pressure phases that probably have cages of reduced symmetry, respectively, so that electron-rattler interactions that govern the superconducting and transport properties of β-RbOs$_2$O$_6$ are significantly affected.

## 1. Introduction

The β-pyrochlore oxides AOs$_2$O$_6$ have generated much attention in research on transition-metal oxides,[1-7] as they exhibit superconductivity and an unusual, local atomic vibration called rattling.[8,9] Rattling is essentially an anharmonic vibration of a heavy ion confined in an oversized atomic cage. In β-pyrochlore oxides, the A atom is located in the $T_d$ site symmetry and surrounded by 6 nearest-neighbor and 12 next-nearest-neighbor oxide atoms, while OsO$_6$ octahedra are connected to each other by vertices to form a three-dimensional skeleton.[10] A virtual size mismatch between the guest A ion and the cage made of the octahedra allows the guest ion to move almost freely with an unusually large atomic excursion in an anharmonic potential inside the cage.[11] Since the mismatch becomes large from Cs to K with decreasing ionic radius of the A ions, the intensity of rattling is enhanced accordingly towards $KOs_2O_6$. Evidence of rattling in β-pyrochlore oxides has been obtained from structural analyses showing large atomic displacement parameters[10,12] or from heat capacity and spectroscopic measurements that showed Einstein-like modes with low energies of 2-7 meV.[5,6,13-17]

The rattling apparently affects the electronic properties of β-pyrochlores. A high resistivity and its anomalous temperature dependence, shown by a concave-downward curvature in a wide temperature range, have been observed and ascribed to a strong scattering of electrons by rattling.[18] Moreover, the increase observed in the spin-lattice relaxation rate of A-nucleus NMR arises from a strong electron-lattice coupling of the same origin.[18,19] Because of this large electron-rattler interaction, β-pyrochore oxides undergo superconducting transitions at relatively high temperatures of $T_c$ = 9.6, 6.3, and 3.3 K for A = K, Rb, and Cs, respectively. Nagao *et al.* suggested that superconductivity is induced by the rattling itself, because the estimated average frequency of phonons mediating Cooper pairing coincides with the energy of rattling for each of the three compounds.[7] Hattori and Tsunetsugu have investigated the role of rattling in the mechanism of superconductivity in the framework of a strong coupling theory of superconductivity and successfully reproduced the observed $T_c$s.[20]

Unique chemical trends for various parameters are observed in the series of β-pyrochlore oxides: the rattling intensity, the magnitude of electron-rattler interactions, and $T_c$ increase systematically from Cs to K.[21] Of particular interest is the fact that the superconductivity changes its character from weak coupling to extremely strong coupling toward K.[6] Since the size of the cage remains almost the same among the three compounds,[10] these variations should be ascribed to the increase in the guest-free space, as the ionic radius of the A ion decreases markedly from Cs to K. Thus, the guest-free space is a key parameter for adjusting the rattling intensity. One experimental method of tuning the guest-free space systematically is to chemically mix two A elements in a crystal. However, this must cause a certain randomness that might mask intrinsic properties. In contrast, squeezing the compound under high pressure would give a better opportunity to study the relationship between the guest-free space and the rattling or electronic properties of β-pyrochlore oxides.

A few high-pressure (HP) experiments have already been carried out using polycrystalline samples of β-pyrochlore oxides.[22-24] Muramatsu *et al.* measured resistivity in a cubic-anvil cell filled with Fluorinert under HP up to 12 GPa and found that, common to the three compounds, $T_c$ initially increases with pressure, saturates, and then decreases to vanish above a critical pressure, resulting in a domelike pressure dependence of $T_c$;[23] the critical pressures were approximately 6, 7, and 12 GPa for K, Rb, and Cs, respectively. On the other hand, Miyoshi *et al.* found, in their magnetization measurements using a diamond-anvil cell (DAC) with Daphne 7373 oil under HPs of up to 10 GPa, similar $T_c$ domes for K and Rb, but a saturating behavior at 8.8 K for Cs.[24] Electronic structure calculations have shown that the density of states decreases gradually and only slightly with increasing pressure.[25] Therefore, the observed complicated pressure dependences of $T_c$ are not understandable in the framework of the simple BCS theory and have not yet been explained satisfactorily.

Note, however, that experimental results can be markedly different between polycrystalline and single-crystal samples in the case of β-pyrochlore oxides.[7] To collect reliable data on the pressure dependence of $T_c$, further HP experimentation using a single crystal is necessary. Recently, Ogusu *et al.* have carried out resistivity measurements under HPs of up to 5 GPa in a high-quality single crystal of $KOs_2O_6$ and found a sudden drop in $T_c$ from 6.5 to 3.3 K at a pressure of 3.6 GPa.[26,27] The sudden drop has been ascribed to a structural transition, by which an enhancement in $T_c$ due to a strong electron-rattler interaction present in the low-pressure cubic phase is abrogated as the rattling of the K ion is completely suppressed or weakened in the high-pressure phase of reduced symmetry.[27] Moreover, two anomalies were observed in the temperature dependence of resistivity in the low-pressure phase of $KOs_2O_6$, which may be due to subtle changes in the crystal structure and thus in the rattling vibration.[27] More recently, Isono and coworkers have carried out HP heat capacity measurements in a DAC filled with Ar on single crystals of $KOs_2O_6$, $RbOs_2O_6$, and $CsOs_2O_6$ and observed abrupt drops in $T_c$ at 5.2, 5.8, and 10.5 GPa, respectively.[28-31] Interestingly, they show that these drops in $T_c$ take place at the same lattice volume of approximately 0.988 $nm^3$ under HP for all the compounds, in which similar structural transitions to those observed in $KOs_2O_6$ may occur.[27]

In this study, we carried out resistivity measurements on a high-quality single crystal of $RbOs_2O_6$ under HPs of up to 6 GPa to investigate the effects of pressure on the superconductivity in more detail and the rattling in the β-pyrochlore oxides. A reliable pressure dependence of $T_c$, which is substantially different from those previously reported for polycrystalline samples, is obtained. We observe a characteristic pressure range between 3.5 and 4.8 GPa, where $T_c$ remains almost constant at 8.8 ± 0.1 K, as well as a sudden drop to 6.3 K at a critical pressure of 4.9 GPa, similarly to that observed in previous resistivity measurements on $KOs_2O_6$. There are probably two structural transitions across which the rattling vibrations of the Rb ion change substantially, affecting the electron-rattler interactions.

## 2. Experimental

Single crystals of $RbOs_2O_6$ were prepared by the chemical transport method in a quartz ampoule at 748 K for 24 h, as reported previously.[7] Two crystals (A and B) were selected and subjected to resistivity measurements by the four-probe method. Most data shown in the present study is from crystal

A of 0.2 × 0.4 × 0.5 mm³ size shown in Fig. 1; essentially the same results were obtained from crystal B. HPs from 1.0 to 6.2 GPa were applied to the sample during the measurements in a cubic-anvil-press apparatus with three pairs of counteranvils made of sintered diamond.[32] Daphne 7474 oil was used as a pressure-transmitting medium; it is liquid below 3.6 GPa at room temperature and has good hydrostatic compression.[33] Isobaric measurements at selected pressure, enabled by applying a constant load on the anvil cell, were performed upon cooling and then heating between 3 and 300 K. Pressure for the next run was increased at room temperature, where the pressure medium was a liquid or sufficiently soft material to generate a uniform, hydrostatic pressure around the sample. The actual pressure exerted on the sample was estimated by measuring changes in resistivity associated with the structural phase transitions of Bi at 2.55, 2.7, and 7.7 GPa, of Te at 4.0 GPa, and of Sn at 9.4 GPa in different runs.[32]

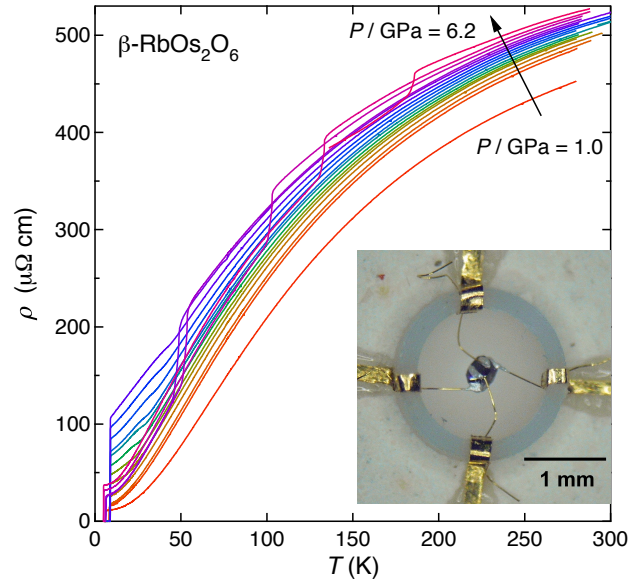

Fig. 1. (Color online) Isobaric resistivities of a single crystal of β-RbOs$_2$O$_6$ measured upon heating after cooling to 3 K under various pressures. The applied pressures were 1.0, 2.6, 3.0, 3.3, 3.6, 3.7, 3.8, 4.0, 4.2, 4.4, 4.6, 4.8, 5.0, 5.2, 5.5, 5.8, and 6.2 GPa from bottom to top at approximately 250 K. The inset shows a photograph of the crystal (crystal A) of 0.2 × 0.4 × 0.5 mm³ size attached with four gold wires to the electrodes. The crystal was placed in a cylindrical Teflon cell filled with Daphne 7474 oil. The cell was put in a cube made of MgO and was hydrostatically compressed by three pairs of counteranvils made of sintered diamond.

## 3. Results

Figure 1 shows seventeen sets of isobaric resistivities $\rho$ measured between $P$ = 1.0 and 6.2 GPa on a single crystal of RbOs$_2$O$_6$. The $\rho$ at 1.0 GPa resembles that reported on a different crystal at ambient pressure,[7] showing a similar concave-downward curvature at high temperatures above ~50 K and a $T^2$ behavior down to the lowest temperature. The residual resistivity ratio is approximately 40 at 1.0 GPa. The overall resistivity curves shift upward with increasing pressure, as observed in KOs$_2$O$_6$.[27] In addition to sharp drops below 10 K owing to superconducting transitions, two more anomalies are observed above 3.6 GPa, which were not detected in previous experiments using a polycrystalline sample. The three anomalies, which are a superconducting transition at $T_c$, a subtle anomaly at $T_o$, and a pronounced anomaly with a large thermal hysteresis at $T_s$, will be described in sequence below.

*3.1 Superconducting transition*

Figure 2 shows the pressure dependence of superconducting transitions. A sharp drop in $\rho$ is observed at most pressures within a transition width less than 0.1 K; it is as sharp as that observed at ambient pressure.[7] A broad superconducting transition is often observed in HP experiments owing to certain inhomogeneity in a sample or in a pressure distribution inside a HP cell. This was in fact the case for previous HP experiments on RbOs$_2$O$_6$, where the transition width became large, e.g., 1.5 and 4 K at 2 and 4 GPa, respectively.[23] Thus, the sharp transitions observed in the present study indicate better sample quality and a more uniform pressure distribution. $T_c$, defined here as a zero-resistive temperature, increases gradually from 6.0 K at 1.0 GPa to 8.7 K at 3.3 GPa. Surprisingly, $T_c$ remains almost the same at 8.8 ± 0.1 K in a wide pressure range from $P_o$ = 3.5 to 4.8 GPa. In contrast, the normal state resistivity $\rho_n$ just above $T_c$ increases markedly by a factor of four in the same pressure range, which confirms a systematic increase in pressure. Other dramatic changes in $T_c$ and $\rho_n$ take place by simply increasing pressure by 0.2 GPa from 4.8 to 5.0 GPa; $T_c$ suddenly drops by 2.5 K to 6.3 K and $\rho_n$ is reduced by one-quarter. We call this critical pressure $P_s$ (~ 4.9 GPa). Further increase in pressure gradually decreases $T_c$ to 4.9 K at 5.8 GPa and slightly increases $\rho_n$. Note that the transition width is always less than 0.1 K, except for two datasets at 5.0 and 5.2 GPa having a width of 0.4 K, immediately after the sudden drop in $T_c$. This broadening may be due to the first-order nature of a transition at $P_s$, as will be described later.

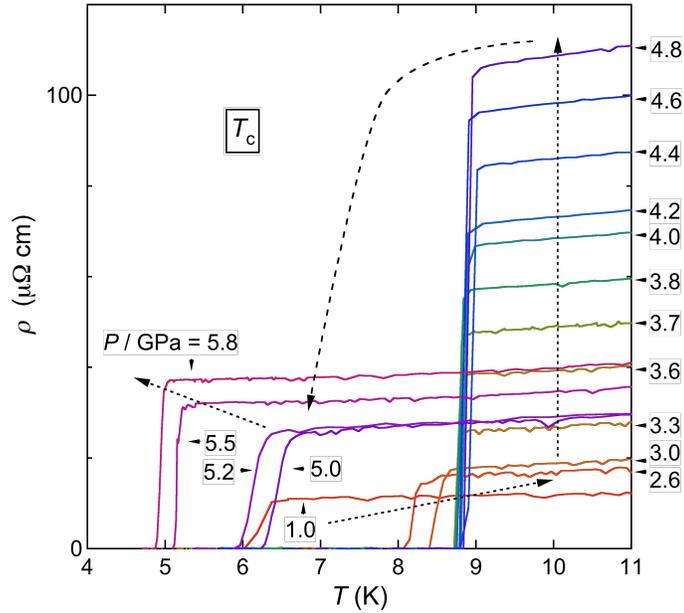

Fig. 2. (Color online) Isobaric resistivities at low temperatures showing evolution of superconducting transitions with pressure. A sequence of data with increasing pressure is shown by broken arrows.

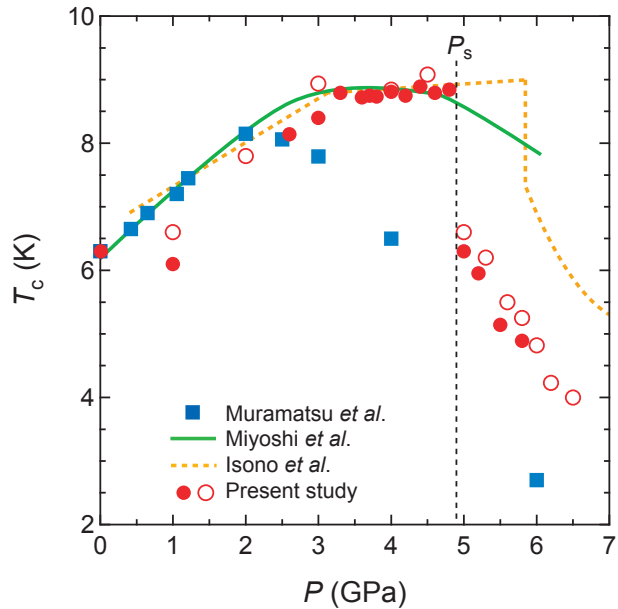

Fig. 3. (Color online) Pressure dependence of $T_c$. The present data from crystals A (filled circle) and B (open circle) are compared with previous data given by Muramatsu et al. (square)[23], Miyoshi et al. (green solid line),[24] and Isono et al. (orange broken line).[28] A sudden drop in $T_c$ is observed at $P_s$ = 4.9 GPa in the present study.

The present characteristic pressure dependences of $T_c$ and $\rho_n$ have not been observed in our previous study. Figure 3 shows a comparison of the pressure dependences of $T_c$ in the present study for crystals A and B with those obtained previously.[23,24,28,29] Muramatsu *et al.* reported a domelike pressure dependence of $T_c$ in their resistivity measurements in a polycrystalline sample. However, it is clearly due to broad transitions in resistivity;[23] $T_c$ was defined at the midpoint of the broad transitions. On the other hand, magnetization measurements by Miyoshi *et al.* are in line with our observations up to $P_s$, but failed to detect a sudden drop in $T_c$, probably because a diamagnetic response was already obscured at approximately $P_s$.[24] Thus, our observation of the sudden drop in $T_c$ at $P_s$ is not contradictory with the previous data and has been made possible owing to the improved sample quality as well as improved experimental conditions. In contrast, the pressure dependence of $T_c$ reported by Isono *et al.* in their heat capacity measurements in a single crystal of $RbOs_2O_6$ is substantially the same as ours, except for the values of $P_s$; their $T_c$ increases gradually with pressure, remains almost constant at 9.3 K above 3 GPa, and suddenly drops to 7 K at 5.8 GPa.[28,29] This confirms that our resistivity results essentially reflect the bulk nature, related to neither a surface event nor a filamentary path, as heat capacity probes bulk property. We do not know, however, the reason for the difference in $P_s$, 4.9 and 5.8 GPa. One speculates that the transition can be seriously affected by the nature of pressure that depends on the equipment used and, more importantly, on the choice of the pressure medium. It is known that some pressure-induced transitions are very sensitive to the choice of the pressure medium. For example, the $\alpha$-to-$\omega$ transition in titanium metal changes from 4.9 to 10.5 GPa depending on the type of pressure medium.[34] Note that there is also a difference in $P_s$ for $KOs_2O_6$ between our experiments and Isono's: $P_s$ = 3.6 and 5.2 GPa, respectively. Although the critical pressures are slightly different, the two experiments should have shown the same phenomenon.

*3.2 Anomalies at high temperatures*

Two anomalies are observed at high temperatures. Figure 4 shows selected sets of $\rho$ below 5.2 GPa in the intermediate temperature range below 60 K. No anomaly is discernible in the resistivity curves below 3.3 GPa, keeping a similar shape and shifting upward gradually with increasing pressure. In contrast, a small upturn emerges in the 3.5 GPa data: $\rho$ begins to shift upward below 20 K from a curve expected by high-temperature extrapolation. We call this temperature $T_o$. As pressure increases, the anomaly moves to higher temperatures, reaching 43 K at 4.8 GPa, and then at 5.0 GPa, the anomaly seems to be replaced by a large drop starting at 48 K. The magnitude of the drop is large, almost 40% of the high-temperature value. We call this temperature $T_s$. These two anomalies have never been observed in previous resistivity measurements of polycrystalline samples.[23]

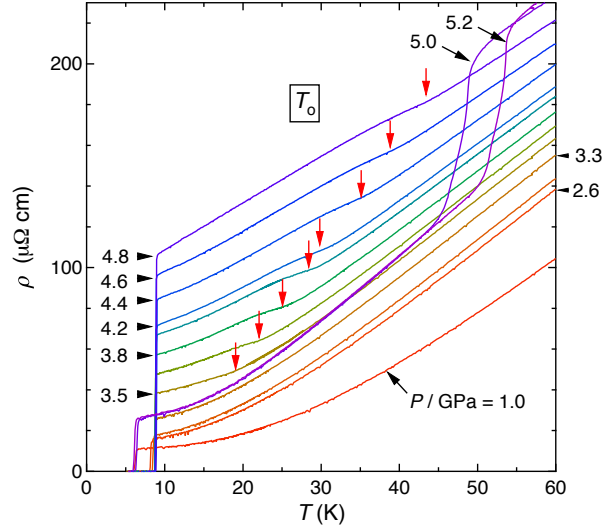

Fig. 4. (Color online) Isobaric resistivities at $P \leq 5.2$ GPa below 60 K. An anomaly appears at $T_o$ above $P_o = 3.5$ GPa up to 4.8 GPa and is replaced by a large drop at $T_s$ above $P_s = 5.0$ GPa.

Figure 5 shows the second anomaly in a wide temperature range. The sharp drop at $T_s$ moves to higher temperatures with increasing pressure. For example, $T_s$ reaches 185 K at a maximum pressure of 6.2 GPa. Note that there is a distinct thermal hysteresis between the cooling and heating curves, as shown in the inset; the difference is 3 K at 5.0 GPa. This is clear evidence of the first-order nature of the transition. Two similar anomalies were observed for crystal B at nearly equal temperatures and pressures. However, they were less pronounced than those for crystal A. It is likely that these transitions are sensitive to the quality of crystals or a slight difference in the experimental setup.

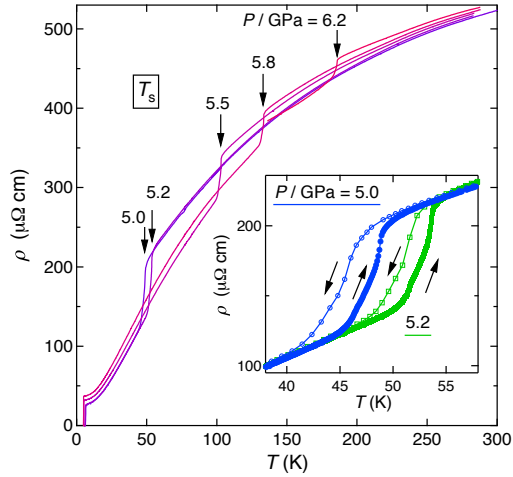

Fig. 5. (Color online) Isobaric resistivities for $P \geq 5.0$ GPa showing a large drop at $T_s$, which moves to higher temperatures with increasing pressure. The inset expands the drops for $P$ = 5.0 and 5.2 GPa, where thermal hysteresis is clearly observed between the heating and cooling curves.

*3.3 P-T phase diagram*

Figure 6 shows a summary of the above results in a *P-T* phase diagram for RbOs$_2$O$_6$. As pressure increases, $T_c$ increases gradually from 6.3 K at ambient pressure to 8.8 K at 3.5 GPa, remains almost constant at 8.8 ± 0.1 K in a wide pressure range between 3.5 ($P_o$) and 4.8 GPa, and suddenly drops to 6.3 K at $P_s$ = 4.9 GPa, followed by a gradual decrease with further pressure increase [see also Fig. 8(a)]. Above $P_o$, a weak anomaly at $T_o$ in $\rho$ appears and moves to higher temperatures with increasing pressure; it is replaced by another large anomaly with a thermal hysteresis at $T_s$ above $P_s$. Note that $P_o$ and $P_s$ decided from the pressure dependence of $T_c$ lie exactly along extrapolations from the $T_o$ and $T_s$ curves in the phase diagram, respectively. Therefore, the observed anomalous pressure dependence of $T_c$ must come from the high-temperature phase transitions, probably of structural origin. We call the high-temperature cubic phase of space group *Fd-3m* as phase I, the intermediate one below the $T_o$ line as phases II, and the HP one as phase III.

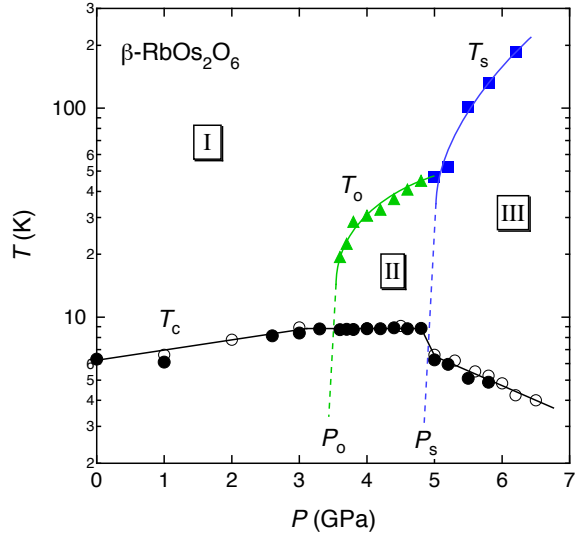

Fig. 6. (Color online) Pressure-temperature phase diagram for β-RbOs$_2$O$_6$. Phase I is a high-temperature, low-pressure phase crystallizing in the cubic pyrochlore structure of the space group *Fd-3m*, where an intense, on-center rattling of the Rb ion is observed. Phase III appears below $T_s$ and above $P_s$, and possesses a collapsed structure with minimal or no rattling. Intermediate phase II below $T_o$ and at $P_o < P < P_s$ may have a slightly distorted structure with off-center rattling. As pressure increases, $T_c$ increases in phase I, remains almost constant in phase II, and suddenly decreases at $P_s$ and then gradually decreases in phase III.

The phase diagram of RbOs$_2$O$_6$ shares a common feature with that of KOs$_2$O$_6$.[27,29] In KOs$_2$O$_6$, $T_c$ increases slightly from 9.60 K at ambient pressure to 9.75 K at 1.0 GPa and then gradually decreases with pressure.[27] Above 1.3 GPa, two anomalies, $T_{o1}$ and $T_{o2}$, appear and move to higher temperatures with increasing pressure. $T_c$ seems to be unaffected by these anomalies, but decreases gradually with pressure. On the other hand, $T_c$ suddenly drops from 6.6 to 3.3 K at $P_s$, and the two anomalies disappear. A third anomaly at $T_s$ shows up above $P_s$. Thus, the two phase diagrams are similar to each other, except for the presence of the two phase transitions for KOs$_2$O$_6$ instead of one transition for RbOs$_2$O$_6$ in intermediate-pressure regions. Powder X-ray diffraction experiments on KOs$_2$O$_6$ revealed a structural transition from cubic to monoclinic or triclinic at approximately $P_s$.[27] Very recently, a similar structural study was performed for RbOs$_2$O$_6$ and also revealed a structural transition near $P_s$.[31] Thus, phase III for either the K or Rb compound must be a collapsed phase with reduced symmetry of the same origin: the rattling of the K/Rb ion is completely suppressed or weakened, so that an enhancement in $T_c$ owing to a strong electron-rattler interaction present in phase I is abrogated.

*3.4 Temperature dependence of resistivity*

The three phases of RbOs$_2$O$_6$ are clearly discriminated from each other by the temperature dependence of the normal state resistivity at low temperatures. Figure 7(a) shows, for phase I, $T^2$ behavior with coefficient $A$ increasing linearly with pressure, as shown in Fig. 8(b). Moreover, the range of temperatures showing the $T^2$ dependence tends to expand to higher temperatures: below ~10 K at 1.0 GPa and ~15 K at 3.3 GPa. In contrast, as shown in Fig. 7(b), $\rho$ becomes proportional to $T$ below $T_o$ for phase II. The temperature range expands to higher temperatures with increasing $P$. In particular, at 4.8 GPa, close to the phase boundary between phases II and III, $\rho$ is perfectly linear below 30 K. On the other hand, in phase III, $\rho$ is proportional to $T^3$ below ~15 K. These marked changes in the temperature dependence demonstrate substantial changes in electron scattering owing to rattling.

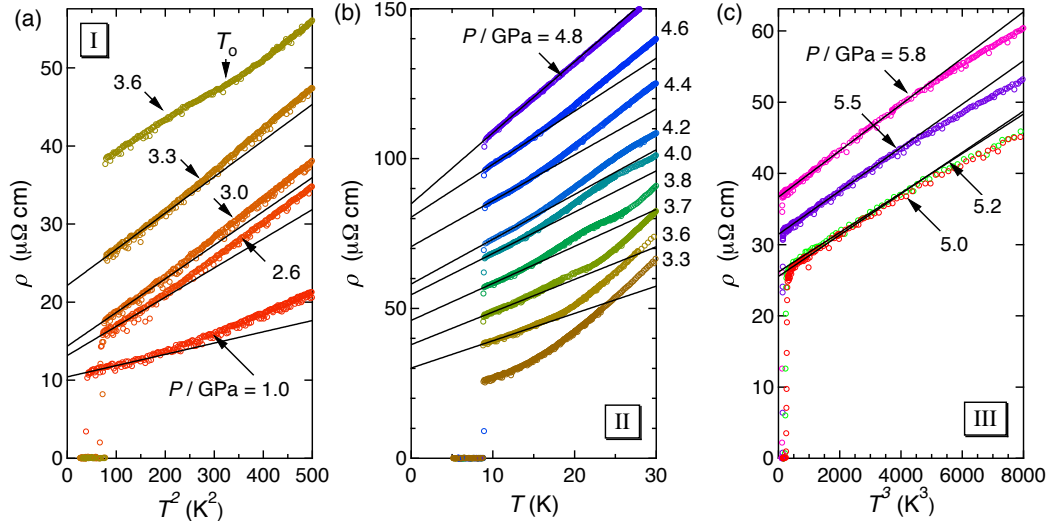

Fig. 7. (Color online) Temperature dependences of resistivity at low temperatures for phases I (a), II (b), and III (c). They are plotted against $T^2$, $T$, and $T^3$, respectively. The line on each curve is a guide for the eye.

To extract more quantitative information on the variations in the temperature dependence of $\rho$, we have fitted the $\rho$ curves below 13 K to the form $\rho = \rho_0 + aT^n$ and plotted the pressure dependence of these parameters in Fig. 8. In phase I, the power $n$ is almost constant at 2. Provided that $n = 2$, the coefficient $A(a)$ increases linearly by a factor of three with pressure. On the other hand, the power decreases and approaches 1 toward the II-III phase boundary in phase II; the large scatter in the data may be due to the limited temperature ranges above $T_c$ for the fitting. Note that the residual resistivity $\rho_0$ remains almost constant in phase I, whereas it markedly increases with pressure and reaches a one-order larger value near the II-III phase boundary in phase II. Then, above $P_s$, $\rho_0$ suddenly recovers, and $n$ jumps to approximately 3.

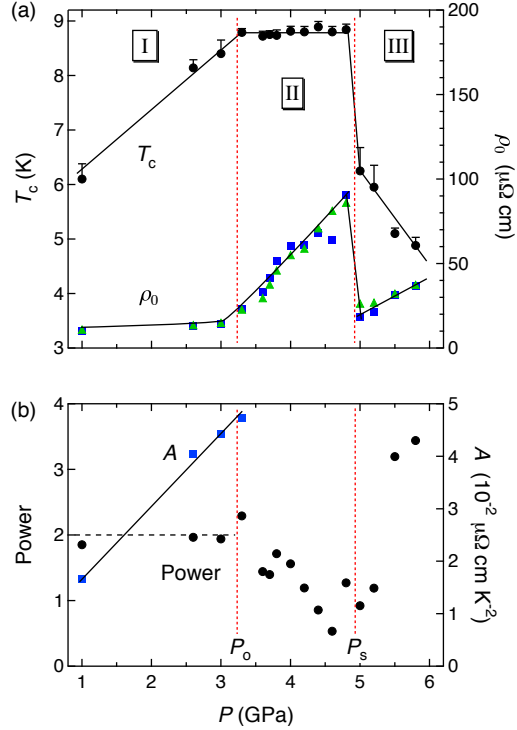

Fig. 8. (Color online) Pressure dependences of $T_c$ and residual resistivity $\rho_0$ (a) and coefficient of $T^2$ term $A$ and power $n$ (b) from fit to $\rho = \rho_0 + aT^n$ below 13 K. The dot for $T_c$ represents a zero-resistivity temperature, and the error bar represents the transition width. The two sets of data for $\rho_0$ are from extrapolations shown in Fig. 7 (triangles) and the fit to the power form (squares).

## 4. Discussion

The present resistivity measurements reveal the presence of two phase transitions under high pressure for $RbOs_2O_6$. It has been well established that, in β-pyrochlores, the magnitude and temperature dependence of resistivity are dominated by electron scattering associated with rattling. Thus, any transition that strongly affects resistivity should be related to a change in the dynamics of rattling vibrations. The observed two transitions at $T_o$ and $T_s$ for $RbOs_2O_6$ must be ascribed to such changes in the rattling of the Rb ion.

In phase I, $\rho$ always shows a $T^2$ behavior with coefficient $A$ increasing with pressure, as shown in Fig. 8(a), indicating the enhancement of the electron-rattler interaction. This is consistent with the findings of a recent HP heat capacity measurement by Iguchi *et al.*: both the effective mass and the degree of jump in heat capacity at $T_c$ increase monotonically with increasing pressure up to $P_s$.[30] Therefore, the increase in $T_c$ is due to this increase. One may think that pressure squeezes the cage, so that the rattling intensity or the anharmonicity of the potential inside the cage tends to be weakened. However, this may not be the case, because the shrinkage of the hard cage made of Os-O bonds is too small in the present pressure range compared with the guest-free-space already existing at ambient pressure, to suppress rattling.[27] Alternatively, compression gives rise to stronger interactions between electrons on the cage and rattlers inside the cage.

Phase III, in analogy to that of $KOs_2O_6$, must have a collapsed structure without any intense rattling: the rattling of the Rb ion is completely suppressed or weakened, so that the increase in $T_c$ owing to strong electron-rattler interactions in phase I is abrogated. The absence of rattling is clearly demonstrated by $T^3$ resistivity instead of $T^2$ resistivity. The origin of the $T^3$ dependence is not clear. We note that a similar $T^3$ dependence has been observed in the α-pyrochlore oxide $Cd_2Re_2O_7$.[35] The $T^3$ resistivity may be related to the scattering of carriers owing to low-energy, dispersive phonons in the common structural framework of these pyrochlore oxides.

In contrast, phase II is anomalous in some aspects. The observed $T$-linear resistivity over a wide temperatue range is quite unusual for phonon scattering. Such a $T$-linear resistivity has been observed in some $f$ electron compounds lying close to an antiferromagnetic quantum critical point, where electrons are scattered by a large spin fluctuation.[36] However, this is absolutely not the case for $RbOs_2O_6$ that has no strong electron correlations.[7] Alternative low-energy fluctuations must be responsible for the $T$-linear dependence, which should also give rise to the large increase in residual resistivity with pressure observed in phase II [Fig. 8(a)]. Another mystery regarding phase II is the insensitivity of $T_c$ to pressure, which may have an important implication for the mechanism of superconductivity but is difficult to understand. There must be a weak structural transition at $T_o$; the anomaly in $\rho$ is small compared with that at $T_s$. Our preliminary structural study provided no evidence of a structural transition. The associated structural change is probably too small to detect by simply modifying the cage slightly.

The high symmetry of the surrounding cage is apparently important for rattling.[20,37] In a cage with low symmetry, a rattler tends to be trapped at one of the off-center positions with a lower potential energy, particularly at low temperatures with weaker thermal vibrations. This is in fact the case for Si-Ge clathrates having low-symmetry cages.[38] The $T_d$ symmetry preserved for the cage in β-pyrochlores must be crucial to the intense rattling of A ions around the on-center position.[20] It is plausible that the transition at $T_o$ breaks the $T_d$ symmetry and modifies the cage slightly, while that at $T_s$ changes the shape of the cage completely.

The possible evolution of the potential for a Rb ion inside a cage is illustrated in Fig. 9. In phase I, a Rb ion vibrates with a large atomic excursion in an anharmonic potential around the center of the cage, which is called on-center rattling. The anharmonicity results in such nonequally spaced energy levels, as shown in Fig. 9(a). The lowest level splitting $\omega_1$ has been estimated to be 60~70 K using heat capacity and spectroscopic measurements.[7,15,17,39] Since the large atomic excursion survives down to low temperatures, strong electron-rattler interactions are generated, which give rise to a large scattering causing the $T^2$ resistivity to appear. The observed increases in coefficient $A$ and the effective mass[30] with pressure mean that the electron-rattler interactions are enhanced, which is the reason why $T_c$ increases with pressure in phase I. In fact, there is a clear tendency, in β-pyrochlore oxides, for $T_c$ and $A$ to increase simultaneously as the anharmonicity increases from the Cs to K compounds.[7]

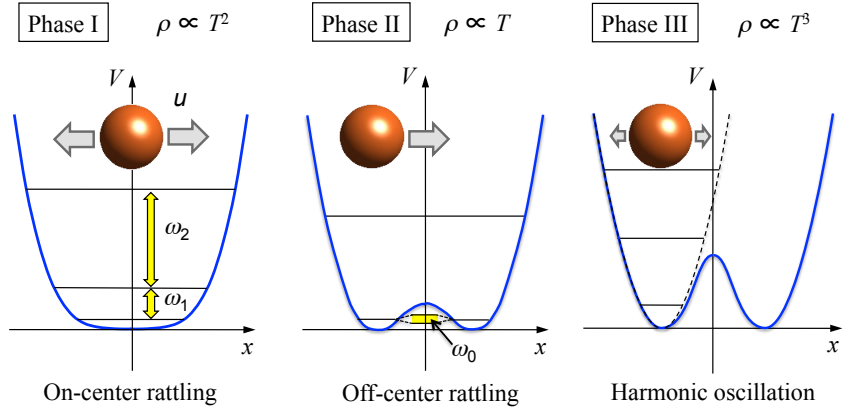

Fig. 9. (Color online) Schematic representation of evolution of rattling vibration of Rb ion. On-center rattling with large anharmonicity occurs for phase I, whereas off-center rattling is expected for phase II. Rattling may be frozen at one of the off-center positions in phase III, resulting in a conventional harmonic oscillation. These changes in rattling give rise to the characteristic temperature dependences of resistivity: $\rho$ is proportional to $T^2$, $T$, and $T^3$ in phases I, II, and III, respectively.

In phase II, in contrast, the $T_d$ symmetry of the cage may be broken as a result of a small structural transition at $T_o$, which produces shallow potential minima at some (maybe four) off-center positions, as shown in Fig. 9(b). Because the potential barrier at the center is still small, rattlers can hop among the off-center positions, which creates a bonding state and an antibonding state separated by a very small energy $\omega_0$. This situation is called off-center rattling. $\omega_0$ must be small compared with $\omega_1$ and can be vanishingly small as the potential barrier is increased. If this extremely low-energy excitation scatters electrons efficiently, the observed $T$-linear resistivity above $T_c$ can be qualitatively explained; the observed temperature range is already high compared with the $\omega_0$ range. Moreover, the increase in $\rho_0$ is predictable, because adding pressure may further distort the cage and increase the barrier height, resulting in a smaller $\omega_0$. The smaller the $\omega_0$, the larger the fluctuations that enhance scattering of carriers as temperature approaches zero. The fact that $\rho$ is slightly enhanced below $T_o$ means that the electron-rattler interactions are enhanced in off-center rattling at phase II than in on-center rattling at phase I.

Finally, at phase III, the cage must be seriously deformed after the first-order transition at $T_s$ and $P_s$. This causes a sudden increase in the barrier height, so that rattlers cannot hop further to the adjacent off-center positions and become confined completely in one of the off-center positions. Thus, there is no more rattling but ordinary harmonic oscillations of a Rb ion, which may cause the temperature dependence in resistivity at a power larger than 2. The observed large reduction in $\rho$ below $T_s$ is evidence of the disappearance of the electron-rattler scattering in phase III. Therefore, what happens under pressure in RbOs$_2$O$_6$ is the evolution of the Rb vibration from on-center rattling in phase I to off-center rattling in phase II, and then to off-center freezing in phase III.

In order to confirm the above scenario, we are carrying out, at present, a structural study of powder samples as well as of a high-quality single crystal of RbOs$_2$O$_6$ under HP. Precise and careful experiments would be required to detect a tiny change in structure at $T_o$. On the other hand, even if this scenario is correct, the constant $T_c$ in phase II remains a mystery. The phonon responsible for the superconductivity at ambient pressure is a rattling phonon with $\omega_1 = 66$ K.[7] The low-energy excitation $\omega_0$ in the off-center rattling is too small to contribute to the superconductivity directly. To understand the constant $T_c$, one must assume a certain compromise between the pressure dependences of some parameters, such as the energy of excitations and the magnitude of electron-rattler interactions or the magnitude of the anharmonicity of rattling. This may not be a mere coincidence, but there must be a means of keeping $T_c$ unchanged.

In the case of KOs$_2$O$_6$, similar changes in the rattling vibration of the K ion must occur. The difference of KOs$_2$O$_6$ from RbOs$_2$O$_6$ is the presence of two phases in the intermediate-pressure region. Probably, owing to the larger anharmonicity for the K ion, symmetry breaking takes place in two steps. In the case of CsOs$_2$O$_6$, we recently carried out similar HP resistivity experiments and found no such intermediate phase but only a collapsed phase with reduced $T_c$ above $P_s \sim 9$ GPa. Off-center rattling may not be realized for a large Cs ion with less anharmonicity.

## 5. Conclusions

We have performed resistivity measurements of a high-quality single crystal of RbOs$_2$O$_6$ under high pressures of up to 6 GPa. A large reduction in $T_c$ from 8.8 to 6.3 K through a first-order phase transition at $P_s$ suggests that the increase in $T_c$ by rattling is abrogated. Another weak transition is observed at an intermediate pressure range above $P_o = 3.5$ GPa and below $P_s$, where resistivity is proportional to $T$, and residual resistivity is enhanced toward the phase boundary at $P_s$. Surprisingly, $T_c$ remains almost constant at $8.8 \pm 0.1$ K in this wide pressure range. To understand these features, it is proposed that the rattling of the Rb ion changes its character from on-center rattling below $P_o$ in phase I, to off-center rattling at $P_o < P < P_s$ in phase II, and then to off-center freezing above $P_s$ in phase III.

## Acknowledgments

We are grateful to K. Hattori for helpful discussion. This work was partly supported by Grant-in-Aids for Scientific Research B (22340092) and Scientific Research on Priority Areas (19052003) provided by MEXT, Japan.